# Improvement of Thermoelectric properties of Lanthanum cobaltate by Sr and Mn co-substitution

Ashutosh Kumar[a,$], D. Sivaprahsam[b] and Ajay D Thakur[a,*]

[a]*Department of Physics, Indian Institute of Technology Patna, Bihta-801 106, INDIA*
[b]*ARC-International, Indian Institute of Technology Madras, Research Park, Chennai-600 133, INDIA*

*\*ajay.thakur@iitp.ac.in, [$]ashutosh.pph13@iitp.ac.in*

We report thermoelectric (TE) properties of Sr and Mn co-substituted $LaCoO_3$ system from room temperature to 700 *K*. Sr-substitutions at La and Mn at Co site in $LaCoO_3$ improves the electrical conductivity ($\sigma$). Thermal conductivity ($\kappa$) of all the samples increases with the increase in temperature but decreases with the substitution in $LaCoO_3$. An estimation of the electronic thermal conductivity ($\kappa_e$) suggests a dominant phonon contribution to thermal conductivity in this system. A maximum value of the figure of merit is 0.14 at 480 K for $La_{0.95}Sr_{0.05}Co_{0.95}Mn_{0.05}O_3$.

**Keywords**: Powders: solid-state reaction, Thermal conductivity, Electrical conductivity, Perovskites.

# 1. Introduction

The tremendous increase in the global energy consumption has a negative environmental baggage associated with exploiting energy resources. There is **a** significant interest to develop conversion technologies which can harvest the waste heat generated during generation, transmission, storage and use of energy [1,2]. Thermoelectric (TE) materials convert heat energy into electrical energy and vice versa *via* TE phenomenon in semiconductors/semi-metals [3,4]. In particular, power generators at a high temperature capable of working in ambient air are highly desirable for harvesting waste heat from vehicle engines and industrial chimneys. For such demands, oxide ceramics are highly promising, but the corresponding TE conversion efficiencies are poor at present.

The figure of merit (ZT) given by $zT=\alpha^2\sigma T/\kappa$ ($\alpha$, $\sigma$, T and $\kappa$ are the Seebeck coefficient, electrical conductivity, absolute temperature and thermal conductivity respectively) is an important parameter which determines the utility of a material for use in TE generators. Thermal conductivity ($\kappa$) consists of charge carrier ($\kappa_e$) and lattice part ($\kappa_{ph}$) [20] i.e, $\kappa=\kappa_e+\kappa_{ph}$. An efficient TE material should have low $\kappa$, high $\alpha$ as well as a high $\sigma$. It is difficult to improve *ZT* as t hese TE parameters are correlated with each other *i.e,* increasing Seebeck coefficient of a material leads to decrease in the values of electrical conductivity. Also, enhancement in electrical conductivity leads to a corresponding increase in the electronic part of thermal conductivity according to Wiedemann-Franz law.

One of the best TE materials, $Bi_2Te_3$, which is commercially available for small-scale energy harvesting is not environmentally attractive due to its toxicity and alternative thermoelectric materials made up of non-toxic elements having a relatively lesser impact on the environment is desirable. Transition metal oxides (TMO) are attractive materials [5-8] to replace $Bi_2Te_3$ since they are less toxic, more stable across multiple thermal cycling and therefore have a weaker environmental impact. Among transition metal oxides, Co-based compounds are of interest due to the possibility of accessing different spin states [9]. The layered cobalt oxide [10], $Na_xCoO_2$, with substitution at Na site has a very high value of $\sigma$, and a low value of $\kappa$ with a large $\alpha$. **The combination of a large $\sigma$, a small $\kappa$, and a large $\alpha$ is required for an effective TE cooling [11,12].** For an effective cooling, $ZT \geq 1$ and has been found in alloys based on bismuth, quantum dot super-lattices and thin-film devices [13, 14]. A large value of Seebeck coefficient ($\alpha$) also occurs in $La_{1-x}Sr_xCoO_3$ up to room temperature [15,16] and the nature of temperature dependence of Seebeck coefficient in this system is very sensitive to Sr concentration as reported elsewhere [17,23-24]. However, most of the research work is focused on single site substitution either at La/Nd site or Co site in rare earth cobaltates [31-33]. The effect of each of the substitution is different and aids in improving the physical parameter required for a good TE materials. The idea of double substitutions was introduced with a view that substitution at La site will increase the electrical conductivity and that at Co site enhances Seebeck coefficient [38]. There have also been attempts to investigate the effect of single-site substitution and co-substitution on TE properties in rare earth manganite systems [34-37]. Studies on thermal transport behavior at high temperatures is worth exploring in samples with co-substitution. In this report, we have investigated the thermal transport behavior of polycrystalline $La_{1-x}Sr_xCo_{1-y}Mn_yO_3$ ($0.00 \leq x \leq 0.10$, $0.00 \leq y \leq 0.10$) samples in the temperature range 300-700 K.

## 2. Experimental

Polycrystalline $La_{1-x}Sr_xCo_{1-y}Mn_yO_3$ ($0.00 \leq x \leq 0.10$, $0.00 \leq y \leq 0.10$) samples were synthesized using standard solid-state reaction technique. Stoichiometric amounts of the precursors $La_2O_3$,

$SrCO_3$, $Co_3O_4$ and $Mn_2O_3$ (Sigma Aldrich, 99.99%) were mixed thoroughly with acetone. The dried mixture was kept in alumina crucibles and calcined at 1273 K for 12 hours in a muffle furnace with 3°/min heating and cooling rate. The calcined powder was ground for 30 minutes to make it more homogenous. Crystallographic structure and phase identification were done using X-ray diffraction (XRD) using a Rigaku TTRX-III diffractometer with Cu-K$_\alpha$ radiation ($\lambda$=1.5406 Å) with a scan rate of 1°/min and a step size of 0.02°. The powder samples were consolidated into sets of pellets of 20 mm diameter each and were sintered at 1373 K for 12 hours with slow heating and cooling rates (2°/min). These pellets were cut into rectangular bars with dimensions 12mm×4mm×4mm using a diamond cutter (IsoMet® Low-Speed Saw) and iso-cut oil (Buehler). They were then ultra-sonicated for 30 minutes to remove any dirt that adhered during cutting of the samples. Rectangular bar-shaped samples were used to measure Seebeck coefficient and electrical resistivity using Seebsys™ (NorECs AS, Norway) with conventional four-probe geometry in the temperature range 300 - 700 K. For measurement of $\alpha$, a temperature difference of 10 K was maintained between both ends of the sample using an auxiliary heater at one end of the sample. Electrical conductivity ($\sigma$) was observed by taking the inverse of the electrical resistivity data in the entire temperature range. The remaining two pellets of 20 *mm* diameter from each set were used for thermal conductivity measurement using non-steady state, transient plane source (TPS) technique which utilizes a sensor element, made of 10 *μm* thick Nickel-metal in the shape of a double spiral [26]. The sensor is sandwiched between the two pellets, in which room temperature thermal conductivity measurements were obtained by supplying 100 mW power for 10 seconds. The room temperature optimized values of parameters including laser power and measurement time were used to measure the high-temperature thermal conductivity of all the samples. The measurement errors for Seebeck coefficient, electrical conductivity, and thermal conductivity were about 3%, however, the corresponding error in the measurement of power factor could be about 10% [39-41].

## 3. Results and discussion

Crystal structure and phase information of all the polycrystalline $La_{1-x}Sr_xCo_{1-y}Mn_yO_3$ ($0.00 \leq x \leq 0.10$, $0.00 \leq y \leq 0.10$) samples have been investigated using XRD followed by Rietveld refinement [18] of all the samples using FullProf$^{TM}$ software. As expected, the XRD pattern of $LaCoO_3$ compound show single phase within the sensitivity of XRD (reference standard data JCPDS 028-1229) indexed as $R\bar{3}c$ space group with reasonable refinement agreement factor ($\chi^2$=1.64) with the characteristic doublet highest intensity peaks at $2\theta = 32.98°$ (104) and 33.40° (110) respectively. Sr and Mn substitutions at La and Co site respectively in $LaCoO_3$ also show the same nature of diffraction pattern as shown in Fig. 1. Substitution at La site leads to shifting of $2\theta$ to lower values while Mn site at Co leads to a shifting to higher values of $2\theta$. Rietveld refinement were done for all the samples and refinement parameters $R_{exp}$, $R_{wp}$, $R_p$, $\chi^2$, $c/a$ ratio, the volume of the unit cell and lattice constants are provided in Table I. Atomic positions with their occupancies are summarized in Table II and the corresponding refinement pattern for $La_{0.95}Sr_{0.05}Co_{0.95}Mn_{0.05}O_3$ is shown in Fig. 2.

The high-temperature thermoelectric properties of Sr and Mn co-substituted $LaCoO_3$ has been studied in the temperature range 300 - 700 K. In Fig. 2 (a), temperature variation of σ is plotted for $La_{1-x}Sr_xCo_{1-y}Mn_yO_3$ ($0.00 \leq x \leq 0.10$, $0.00 \leq y \leq 0.10$) measured using DC four-probe technique. In the temperature range specified, we observe the increase in the values of electrical conductivity with the Sr-substitution at La site and Mn-substitution at Co site. This may be attributed to the increase in carrier concentration (hole in this case) with Sr-and Mn-substitutions. The increase in electrical conductivity with the increase in temperature shows the semiconducting nature of the samples. In order to estimate the activation energy in these compositions, we have made the Arrhenius plot (ln$\sigma$ vs 1/$T$) for the polycrystalline samples of La$_{1-x}$Sr$_x$Co$_{1-y}$Mn$_y$O$_3$ ($0.00 \leq x \leq 0.10$, $0.00 \leq y \leq 0.10$) as shown in Fig. 3. From the plot, we have clearly identified two regions for all the samples and linear fitted data corresponding to these regions are shown in Fig. 3 [27]. The activation energy of all the samples in different regions is obtained from the slope of these linear fitted curves and their values with temperature range are shown in Table III.

In Fig. 2(b), we plot Seebeck coefficients ($\alpha$) as a function of temperature (300 – 700 $K$) for La$_{1-x}$Sr$_x$Co$_{1-y}$Mn$_y$O$_3$ ($0.00 \leq x \leq 0.10$, $0.00 \leq y \leq 0.10$). The positive values of Seebeck coefficient of all the samples indicate the p-type nature of conduction in these oxides. At room temperature, we did not get negative values of Seebeck coefficients which occasionally is the case due to oxygen deficiency for LaCoO$_3$[29]. The Seebeck coefficient ($\alpha$) of La$_{1-x}$Sr$_x$Co$_{1-y}$Mn$_y$O$_3$ ($0.00 \leq x \leq 0.10$, $0.00 \leq y \leq 0.10$) changes from 336 µV/K for LaCoO$_3$ (x=0.00, y=0.00) to 220 µV/K for La$_{0.95}$Sr$_{0.05}$CoO$_3$ (x=0.05, y=0.00) and 258 µV/K for La$_{0.95}$Sr$_{0.05}$Co$_{0.95}$Mn$_{0.05}$O$_3$ (x=0.05, y=0.05). This shows that simultaneous substitution of Sr and Mn at La and Co site respectively enhances the Seebeck coefficient value. Koshibae et al. [9] showed that the strong correlation of 3d electrons and characteristic spin degeneracy in cobalt oxide caused the large value of Seebeck coefficient. We could not obtain a large value of α (~700 µV/K for x=0.05 around 300 K) as reported by Androulakis et al. [15] however the behavior of Seebeck coefficient of Sr-substituted LaCoO$_3$ samples is similar to that reported by Sehlin et al. [23] and Ohtani et al. [28]. It has also been observed that, below 550 K, $\alpha$ decreases with increasing temperature. This remarkable decrease in $\alpha$ with increase in temperature below 550 K is attributed due to the spin state transitions [22,25]. As Sr is substituted at La sites, holes are introduced into the system which suppresses the spin state transition and values of α decrease dramatically. Further Mn-substitution at Co sites leads to an increase in the value of Seebeck coefficient. In the high-temperature region, the Seebeck coefficient of substituted cobaltates can be determined by Heikes formula [9, 17, 21]

$$\alpha = -\left(\frac{k_B}{e}\right) ln\left[\left(\frac{c_3}{c_4}\right)\left(\frac{n}{(1-n)}\right)\right]$$

where, $k_B$ is Boltzmann's constant, $e$ is electronic charge, $n$ is the concentration fraction of $Co^{4+}$ ions, and, $c_3$ and $c_4$ represents respectively the number of possible configuration of $Co^{3+}$ and $Co^{4+}$ [9, 17]. Values of $c_3$ and $c_4$ can be determined using Hund's rule coupling, crystal field splitting energy and temperature. The two fractions $\left(\frac{c_3}{c_4}\right)$ and $\left(\frac{n}{(1-n)}\right)$ determine not only the magnitude but also the sign of α. At high temperatures, the holes generated due to the thermally activated low-spin $Co^{4+}$ from low-spin $Co^{3+}$ saturate and approaches a constant value. Therefore at high temperatures, α tends to be independent of temperature [17] for all the samples of $La_{1-x}Sr_xCo_{1-y}Mn_yO_3$ (0.00 ≤ x ≤ 0.10, 0.00 ≤ y ≤ 0.10) as observed in the top panel of Fig. 3. Also an increase in the substitution content leads to a decrease in Seebeck coefficient for all the substituted samples. It has been observed that hole doping is useful in reducing the electrical resistivity of the samples, however it also leads to a decrease in Seebeck coefficient which shows that these two parameters are strongly correlated [30]. Substitution with other cations Ba [31] and Ca[32] also increase the electrical conductivity and decrease the Seebeck coefficient, however with co-substitution the decrease in Seebeck coefficient is less.

As discussed, total thermal conductivity (κ) consists of phonon thermal conductivity ($κ_{ph}$) and electronic thermal conductivity ($κ_e$), calculated using $κ_e=L_0σT$, where $L_0$ is Lorentz number (1.50× $10^{-8}$ $V^2/K^2$) for non-degenerate semiconductors [42]. The $κ_{ph}$ can be calculated by subtracting the $\kappa_e$ part from total thermal conductivity $\kappa$. In Fig. 3 (a) total thermal conductivity $\kappa$ as a function of temperature for $La_{1-x}Sr_xCo_{1-y}Mn_yO_3$ ($0.00 \leq x \leq 0.10$, $0.00 \leq y \leq 0.10$) is plotted. Decrease in the value of total thermal conductivity ($\kappa$) has been observed at room temperature with Sr and Mn substitution as it enhances the phonon-phonon scattering. Total thermal conductivity has been found to increase with increase in temperature for $La_{1-x}Sr_xCo_{1-y}Mn_yO_3$ which may be due to **a** dramatic increase in the $\kappa_e$ (shown in Fig. 3 (b)). Temperature variation of phonon thermal conductivity ($\kappa_{ph}$) is shown in Fig. 3 (c). This shows that $\kappa_{ph}$ varies differently with parent compound and substituted samples. For $LaCoO_3$, $\kappa_{ph}$ increases almost linearly up to 500 $K$ and above this temperature the $\kappa_{ph}$ starts saturating, this behavior may be due to the fact that at high temperatures where the atomic displacements are large, more number of phonons participate in the collision leading to a decrease in the phonon mean free path and hence reduces the thermal conductivity.

It has been observed that total thermal conductivity $\kappa$ is dominated by phonon thermal conductivity $\kappa_{ph.}$ Although electronic thermal conductivity is enhanced due to high value of electrical conductivity, however phonon contribution of total thermal conductivity is 50 - 70 % of the total thermal conductivity, suggesting that *ZT* can be enhanced by reducing the lattice part of thermal conductivity by either proper substitutions or optimized microstructure to enhance phonon boundary scattering. Based on the values of α σ, and κ, power factor ($\alpha^2\sigma$) and figure of merit (*ZT*) are calculated. In Fig. 4. power factor as a function of temperature is shown and *ZT* as a function of temperature is shown in Fig. 5 for polycrystalline $La_{1-x}Sr_xCo_{1-y}Mn_yO_3$ ($0.00 \leq x \leq 0.10$, $0.00 \leq y \leq 0.10$). A noticeable enhancement in the power factor and figure of merit has been observed for x=0.05, y=0.05 substitution in the $La_{1-x}Sr_xCo_{1-y}Mn_yO_3$ system.

The figure of merit for LaCoO$_3$ increases from the room temperature and reaches a maximum at 560 $K$ ($ZT_{max}$=0.03) and then started decreasing as the thermopower decreases at high temperatures. Similarly $ZT$= 0.082 at 300 K and reaches a maximum at 480 $K$, $ZT_{max}$=0.104 and then decreases for La$_{0.95}$Sr$_{0.05}$CoO$_3$. The maximum value of $ZT_{max}$=0.14 was observed for the sample with x=0.05, y=0.05 at 480 K. It can be seen from the present studies that the Sr and Mn co-substitution leads to an improvement in thermoelectric properties of LaCoO$_3$.

## 4. Conclusions

Sr and Mn co-substituted LaCoO$_3$ polycrystalline samples have been prepared using the standard solid-state method. Structural properties were studied extensively using XRD followed by Rietveld analysis and their thermoelectric properties were measured. The Seebeck coefficients of La$_{1-x}$Sr$_x$Co$_{1-y}$Mn$_y$O$_3$ (0.00 ≤ x ≤ 0.10, 0.00 ≤ y ≤ 0.10) were found to decrease with increasing temperatures in the entire temperature range studied. The increase in electrical conductivity was also observed with increasing temperature and its value reached ~ 700 $S/cm$ (x=0.05, y=0.05) indicating the metallic nature of the samples at high temperatures. The decrease in Seebeck coefficient and an increase in electrical conductivity of all the samples are strongly affected by the spin-state transitions. The total thermal conductivities of these oxides increase with increasing temperature and it has been observed that the phonon thermal conductivity dominates over total thermal conductivity over the entire temperature range studied. The obtained maximum $ZT$ value is 0.14 at 480 $K$ for La$_{0.95}$Sr$_{0.05}$Co$_{0.95}$Mn$_{0.05}$O$_3$.

## Acknowledgements

AK and ADT would like to acknowledge Ministry of Human Resources and Development (MHRD), Government of India for financial support.

**FIGURE CAPTIONS:**

**Fig.1.** (Color Online) X-ray diffraction pattern of $La_{1-x}Sr_xCo_{1-y}Mn_yO_3$ ($0.00 \leq x \leq 0.10$, $0.00 \leq y \leq 0.10$). Indexing of peaks has been done as per ICDD file no. 028-1229.

**Fig. 2.** (Color Online) Rietveld refinement pattern of $La_{0.95}Sr_{0.05}Co_{0.95}Mn_{0.05}O_3$. The refinement parameter and corresponding results are prescribed in Table I & II.

**Fig. 3.** (Color Online) Electrical conductivity ($\sigma$) and Seebeck coefficient ($\alpha$) are plotted as a function of temperature in $La_{1-x}Sr_xCo_{1-y}Mn_yO_3$ ($0.00 \leq x \leq 0.10$, $0.00 \leq y \leq 0.10$).

**Fig. 4.** (Color Online) Arrhenius plot: variation of electrical conductivity ($\ln\sigma$) as a function of inverse of temperature ($1/T$) for the sample of $La_{1-x}Sr_xCo_{1-y}Mn_yO_3$ ($0 \leq x \leq 0.10$, $0 \leq y \leq 0.10$).

**Fig. 5.** (Color Online) (a) Total thermal conductivity ($\kappa$), (b) electronic thermal conductivity ($\kappa_e$) and (c) phonon thermal conductivity ($\kappa_{ph}$) as a function of temperature measured using TPS for $La_{1-x}Sr_xCo_{1-y}Mn_yO_3$ ($0.00 \leq x \leq 0.10$, $0.00 \leq y \leq 0.10$).

**Fig. 6**. (Color Online) Power Factor ($\alpha^2\sigma$) and Figure of merit ($ZT$) as a function of temperature calculated using thermoelectric parameters for $La_{1-x}Sr_xCo_{1-y}Mn_yO_3$ ($0.00 \leq x \leq 0.10$, $0.00 \leq y \leq 0.10$).

**TABLE CAPTIONS:**

**TABLE I.** Refinement parameters $R_{wp}$, $R_{exp}$, $R_p$, $\chi^2$, Lattice parameters $a$, $c$ and $c/a$ ratio of $La_{1-x}Sr_xCo_{1-y}Mn_yO_3$ ($0 \leq x \leq 0.10$, $0 \leq y \leq 0.10$) calculated from the Rietveld refinement of the XRD patterns.

**TABLE II.** Atomic position and their occupancy as a function of Sr and Mn co-substitution in $La_{1-x}Sr_xCo_{1-y}Mn_yO_3$ ($0 \leq x \leq 0.10$, $0 \leq y \leq 0.10$).

**TABLE III.** Activation energy ($E_a$) as a function of Sr and Mn co-substitution in $La_{1-x}Sr_xCo_{1-y}Mn_yO_3$ ($0 \leq x \leq 0.10$, $0 \leq y \leq 0.10$).

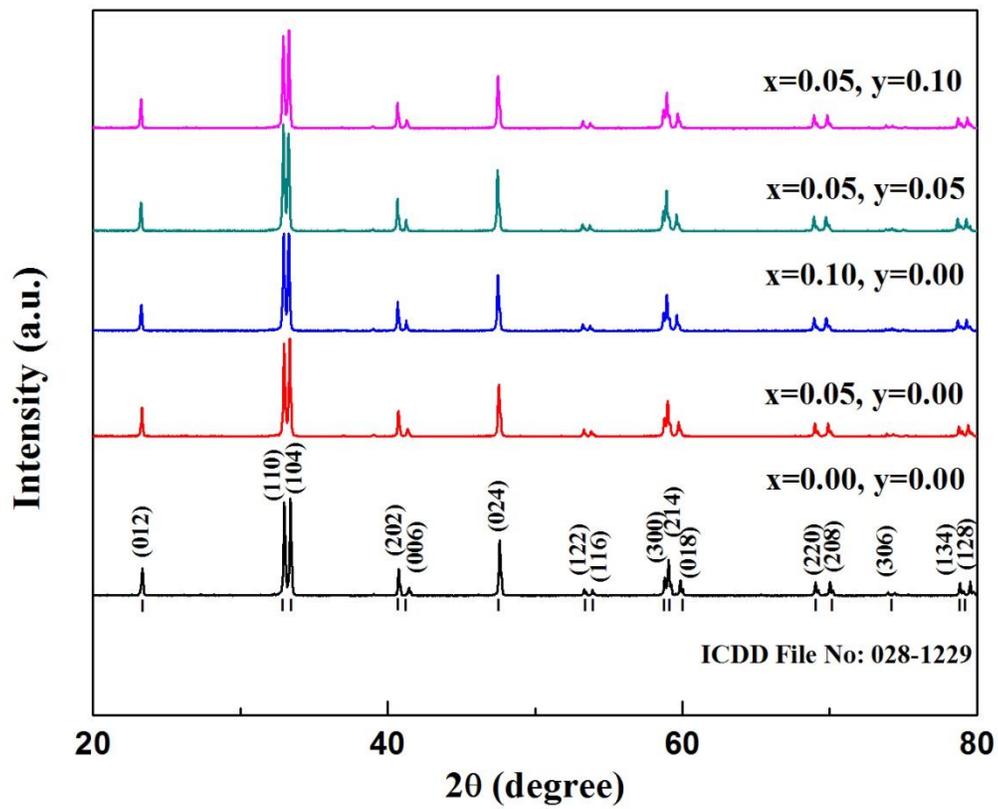

Figure 1

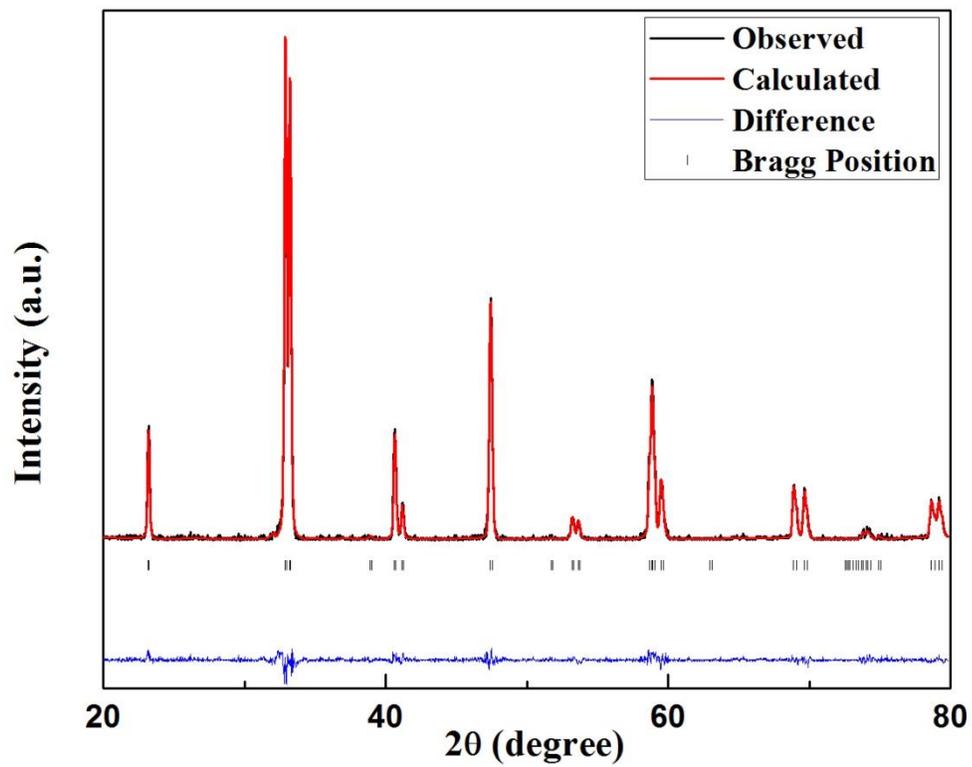

**Figure 2**

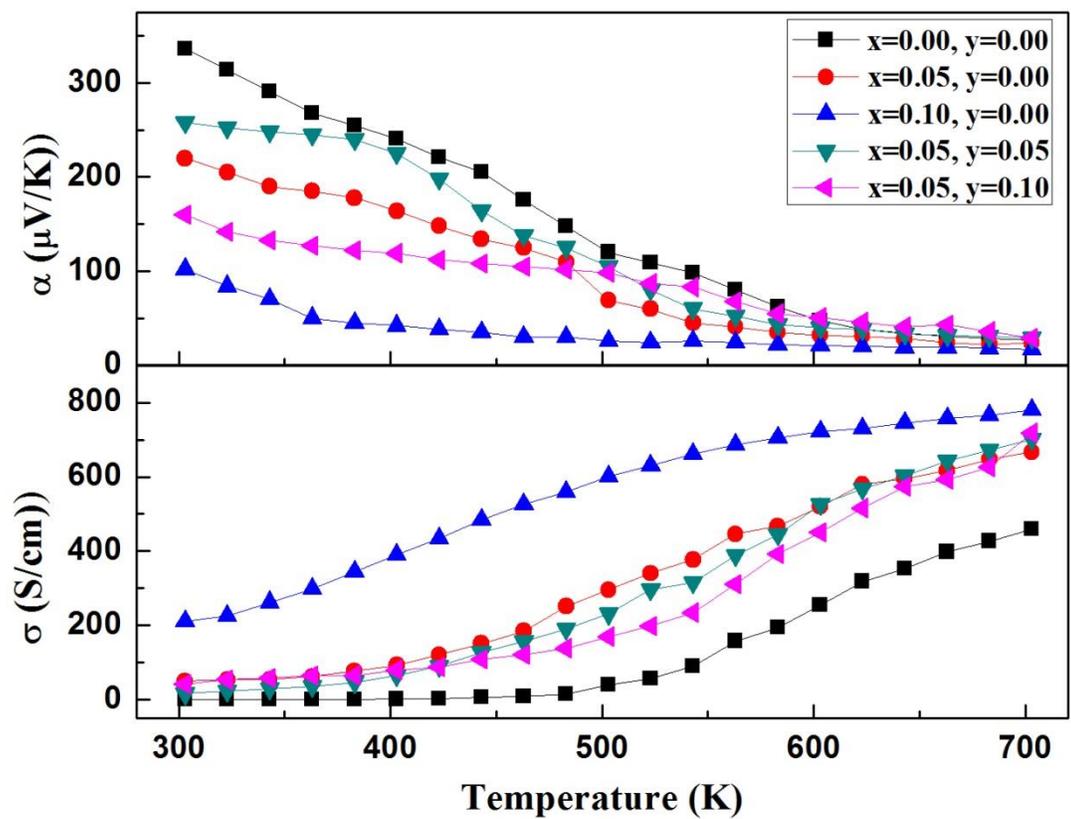

Figure 3

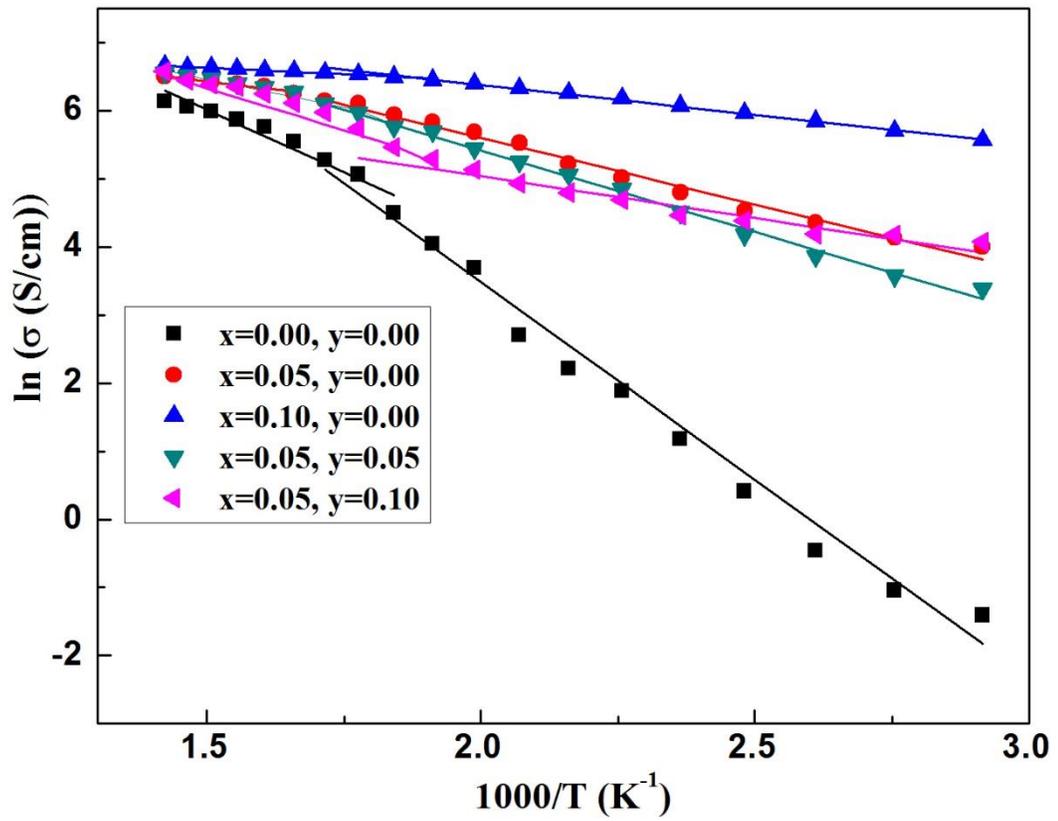

Figure 4

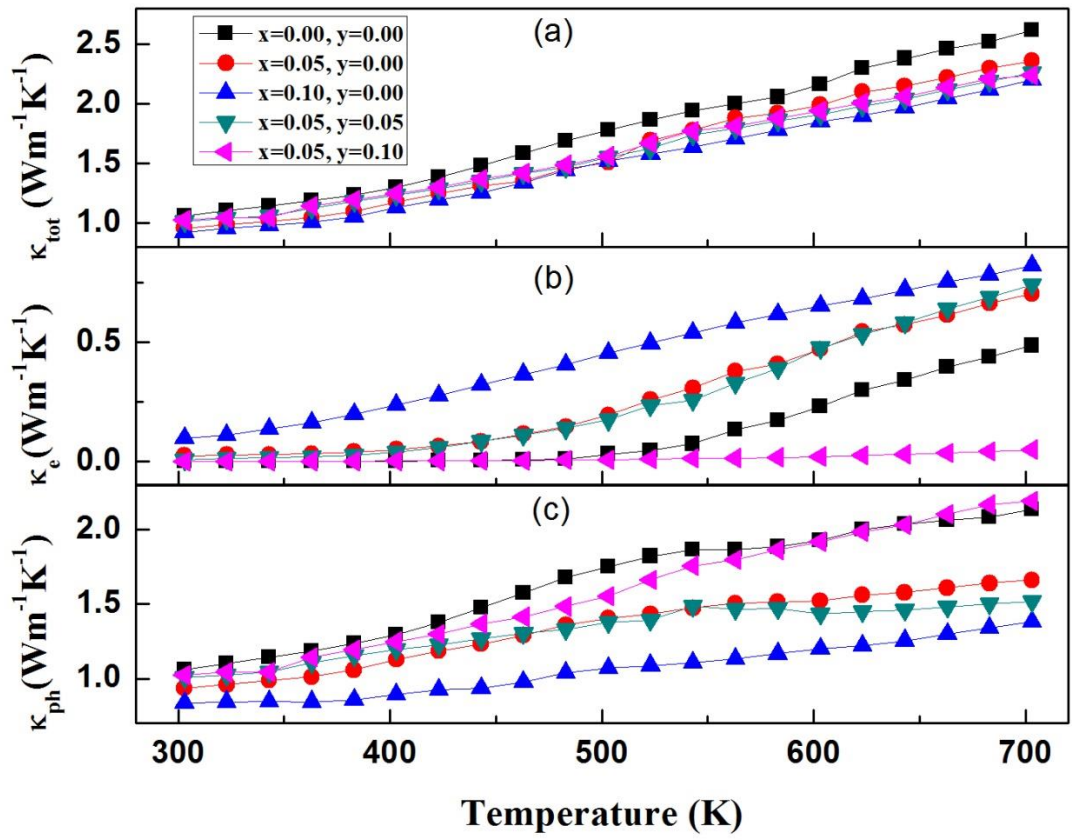

**Figure 5**

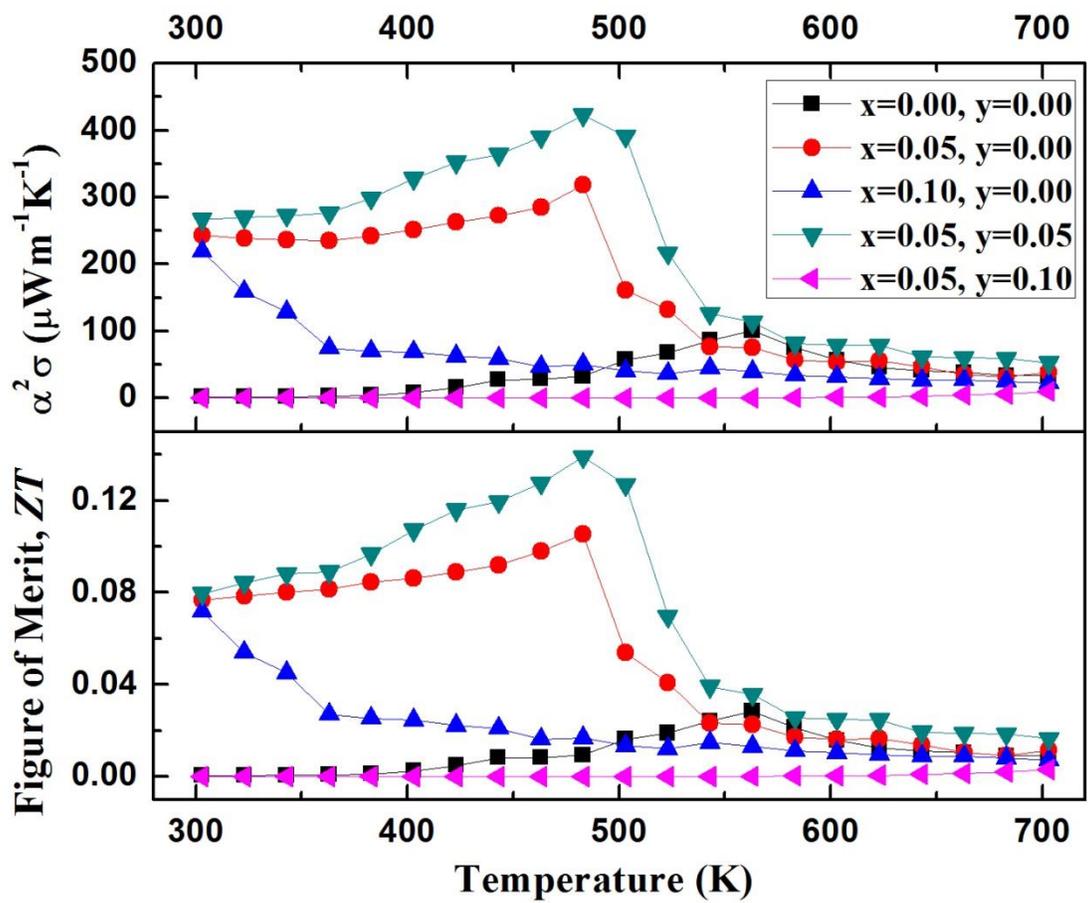

Figure 6

**TABLE I.**

| System Under Study La$_{1-x}$Sr$_x$Co$_{1-y}$Mn$_y$O$_3$ | a (Å) | c (Å) | c/a | V(Å$^3$) | R$_{wp}$ | R$_{exp}$ | R$_p$ | χ$^2$ |
|---|---|---|---|---|---|---|---|---|
| x=0.00, y=0.00 | 5.439(5) | 13.084(6) | 2.405 | 335.28 | 23.3 | 18.19 | 19.3 | 1.64 |
| x=0.05, y=0.00 | 5.440(5) | 13.115(5) | 2.410 | 336.20 | 21.6 | 17.51 | 18.4 | 1.52 |
| x=0.10, y=0.00 | 5.447(3) | 13.155(8) | 2.418 | 338.00 | 23.3 | 18.60 | 14.8 | 1.56 |
| x=0.05, y=0.05 | 5.443(3) | 13.154(8) | 2.416 | 337.48 | 20.8 | 16.84 | 14.7 | 1.53 |
| x=0.05, y=0.10 | 5.446(3) | 13.157(4) | 2.416 | 337.93 | 19.7 | 16.20 | 14.2 | 1.48 |

**TABLE II.**

|  | Site | X | Y | Z | Biso | Occupancy |
|---|---|---|---|---|---|---|
| $R\bar{3}c$ (x=0.00, y=0.00) | | | | | | |
| La | 6a | 0.000 | 0.000 | 0.250 | 0.00 | 0.1655 |
| Co | 6b | 0.000 | 0.000 | 0.000 | 0.00 | 0.1855 |
| O | 18c | 0.453 | 0.000 | 0.250 | 0.00 | 0.5126 |
| $R\bar{3}c$ (x=0.05, y=0.00) | | | | | | |
| La/Sr | 6a | 0.000 | 0.000 | 0.250 | 0.00 | 0.1441/0.0249 |
| Co | 6b | 0.000 | 0.000 | 0.000 | 0.00 | 0.1726 |
| O | 18c | 0.467 | 0.000 | 0.250 | 0.00 | 0.4591 |
| $R\bar{3}c$ (x=0.10, y=0.00) | | | | | | |
| La/Sr | 6a | 0.000 | 0.000 | 0.250 | 0.00 | 0.1424/0.0251 |
| Co | 6b | 0.000 | 0.000 | 0.000 | 0.00 | 0.1712 |
| O | 18c | 0.459 | 0.000 | 0.250 | 0.00 | 0.4951 |
| $R\bar{3}c$ (x=0.05, y=0.05) | | | | | | |
| La/Sr | 6a | 0.000 | 0.000 | 0.250 | 0.00 | 0.1437/0.0241 |
| Co/Mn | 6b | 0.000 | 0.000 | 0.000 | 0.00 | 0.1487/0.0258 |
| O | 18c | 0.461 | 0.000 | 0.250 | 0.00 | 0.4861 |
| $R\bar{3}c$ (x=0.05, y=0.10) | | | | | | |
| La/Sr | 6a | 0.000 | 0.000 | 0.250 | 0.00 | 0.1424/0.0251 |
| Co/Mn | 6b | 0.000 | 0.000 | 0.000 | 0.00 | 0.1387/0.0341 |
| O | 18c | 0.447 | 0.000 | 0.250 | 0.00 | 0.4951 |

**TABLE III.**

| $La_{1-x}Sr_xCo_{1-y}Mn_yO_3$ | Region | Temperature Range (K) | Activation energy (eV) |
|---|---|---|---|
| **x=0.00, y=0.00** | I | 320-560 | 0.49 |
| | II | 560-700 | 0.32 |
| **x=0.05, y=0.00** | I | 320-480 | 0.18 |
| | II | 480-700 | 0.08 |
| **x=0.10, y=0.00** | I | 320-560 | 0.07 |
| | II | 560-700 | 0.03 |
| **x=0.05, y=0.05** | I | 320-480 | 0.21 |
| | II | 480-700 | 0.14 |
| **x=0.05, y=0.10** | I | 320-480 | 0.10 |
| | II | 480-700 | 0.21 |